\documentstyle[11pt]{article}
\makeatletter
\@addtoreset{equation}{section}
\newdimen\normalarrayskip
\newdimen\minarrayskip
\normalarrayskip\baselineskip
\minarrayskip\jot
\newif\ifold \oldtrue \def\new{\oldfalse}
\def\arraymode{\ifold\relax\else\displaystyle\fi}

\def\@arrayskip{\ifold\baselineskip\z@\lineskip\z@
  \else
  \baselineskip\minarrayskip\lineskip2\minarrayskip\fi}
\def\@arrayclassz{\ifcase \@lastchclass \@acolampacol \or
\@ampacol \or \or \or \@addamp \or
 \@acolampacol \or \@firstampfalse \@acol \fi
\edef\@preamble{\@preamble
 \ifcase \@chnum
  \hfil$\relax\arraymode\@sharp$\hfil
  \or $\relax\arraymode\@sharp$\hfil
  \or \hfil$\relax\arraymode\@sharp$\fi}}
\def\@array[#1]#2{\setbox\@arstrutbox=\hbox{\vrule
  height\arraystretch \ht\strutbox
  depth\arraystretch \dp\strutbox
  width\z@}\@mkpream{#2}\edef\@preamble{\halign \noexpand\@halignto
\bgroup \tabskip\z@ \@arstrut \@preamble \tabskip\z@ \cr}%
\let\@startpbox\@@startpbox \let\@endpbox\@@endpbox
 \if #1t\vtop \else \if#1b\vbox \else \vcenter \fi\fi
 \bgroup \let\par\relax
 \let\@sharp##\let\protect\relax
 \@arrayskip\@preamble}
\@addtoreset{equation}{section}
\makeatother

\def\theequation{\thesection.\arabic{equation}}

\def\lvm{\leavevmode\hbox to\parindent{\hfill}}
\def\req#1{(\ref{#1})}

\def\BE{\begin{equation}}
\def\EE{\end{equation} }
\def\BA{\begin{array}} 
\def\EA{\end{array}}

\def\const{\mathop{\rm const}\nolimits}
\def\rank{\mathop{\rm rank}\nolimits}

\def\bar{\overline}
\def\frac#1#2{\mathchoice{{\textstyle{{#1}\over{#2}}}}{{#1\over#2}}%
  {{#1\over#2}}{{#1\over#2}}}

\def\d{\partial}

\def\dpl{\stackrel{\rightarrow}{\d}}
\def\dpr{\stackrel{\leftarrow}{\d}}

\def\dl#1#2{{\stackrel{\rightarrow}{\d}\!#1\over\d #2}}

\def\ddl#1#2{{\stackrel{\rightarrow}{\d}\over\d #2}#1}
\def\ddr#1#2{#1{\stackrel{\leftarrow}{\d}\over\d #2}}

\def\e{\epsilon}

\def\half{\frac{1}{2}}

\def\sign#1#2{(-1)^{(\e(#1)+1)(\e(#2)+1)}}
\def\sig#1#2{(-1)^{\e(#1)\e(#2)}}

\def\cL{{\cal L}}
\def\cM{{\cal M}}

\def\CM{C_{\cM}}

\def\tilde{\widetilde}

\def\PRD{Phys.\ Rev.\ D}
\def\NPB{Nucl.\ Phys.\ B}
\def\PLB{Phys.\ Lett.\ B}
\def\MPLA{Mod.\ Phys.\ Lett.\ A}
\def\CMP{Commun.\ Math.\ Phys.}

\def\JMP{J.\ Math.\ Phys.}

\newtheorem{fact}{Proposition}[section]

\newtheorem{dfn}[fact]{Definition}

\newtheorem{thm}[fact]{Theorem}

\begin{document}
\hfuzz=1pt
\addtolength{\baselineskip}{2pt}

\begin{flushright}
  {\tt hep-th/9708077}
\end{flushright}
\thispagestyle{empty}

\begin{center}
  {\Large{\sc On the Canonical Form of a Pair of Compatible
      Antibrackets}}\\[16pt]
  {\large M.~A.~Grigoriev and A.~M.~Semikhatov}\\[4pt]
  {\small\sl Tamm Theory Division, Lebedev Physics Institute, Russian
    Academy of Sciences}
\end{center}

\addtolength{\baselineskip}{-2pt}

\noindent\hbox to.05\hsize{\hfill}
\parbox{.9\hsize}{\footnotesize
  In the triplectic quantization of general gauge theories, we prove a
  `triplectic' analogue of the Darboux theorem: we show that the
  doublet of compatible antibrackets can be brought to a {\it
    weakly-canonical\/} form provided the general triplectic axioms
  of~\cite{[BMS]} are imposed together with some additional
  requirements that can be formulated in terms of marked functions of
  the antibrackets. The weakly-canonical antibrackets involve an
  obstruction to bringing them to the canonical form.  We also
  classify the `triplectic' odd vectors fields compatible with the
  weakly-canonical antibrackets and construct the Poisson bracket
  associated with the antibrackets and the odd vector fields.  We
  formulate the $Sp(2)$-covariance requirement for the antibrackets
  and the vector fields; whenever the obstruction to the canonical
  form of the antibrackets vanishes, the $Sp(2)$-covariance
  condition implies the canonical form of the triplectic vector fields.}

  \addtolength{\baselineskip}{2pt}

\section{Introduction}\lvm
Triplectic quantization of general gauge
theories~\cite{[BM0],[BMS],[BM]} was formulated as a generalization of
the $Sp(2)$-symmetric Lagrangian quantization~\cite{[Hull],[BLT]},
into which the ghost and antighost fields enter in a symmetric way
(which is in contrast to the standard BV-formalism~\cite{[BV]}). In
the triplectic formalism, one introduces a pair of antibracket
operations (in fact, a pair of odd BV $\Delta$-operators) that are
required to satisfy certain compatibility conditions. In addition,
one introduces two odd vector fields~\cite{[BLT],[BM0],[DN]} which,
again, should agree with the antibrackets in a certain sense.  All
these objects allow one two formulate the triplectic master-equations,
whose solutions enter the corresponding path integral.

The starting point of the triplectic quantization prescription is the
quantization~\cite{[BLT],[Hull]} using the coordinates on the field
space in which field-antifield identifications are explicit and the
antibrackets are written in the `canonical' form. As with the
conventional BV formalism, where the {\it covariant\/} formulation is
now available~\cite{[BT],[HZ],[SZ]}, the aim of the triplectic
formalism is to give a formulation that is covariant with respect to
changing coordinates on the field space. This amounts to replacing the
`triplectic phase space' with a supermanifold $\cM$ whose local
coordinates are not separated into fields and antifields explicitly.
The fundamental objects such as the antibrackets\footnote{In this
  paper, we concentrate on the antibrackets rather than on the BV
  $\Delta$-operators and, thus, do not consider the measure on the
  triplectic manifold.} are then introduced axiomatically, similarly
to how the Poisson bracket is defined on a general Poisson manifold.
However, there is a significant gap in the triplectic formulation,
which can, potentially, be a source of problems in the formalism.
Recall that, for the Poisson manifolds, the Darboux theorem guarantees
the existence of local coordinates in which the Poisson bracket takes
the canonical form. As regards the {\it antisymplectic\/} manifolds
employed in the covariant version of the BV formalism, a similar
theorem is usually assumed~\cite{[SHANDER]}.  Thus, the covariant
formulation is eventually equivalent to the original
formalism~\cite{[BV]} in the field-antifield space.  For the
triplectic objects, on the other hand, Darboux-type theorems are not
known, and the entire construction of~\cite{[BMS],[BM]} relies on the
conjecture that {\it some\/} theorem of this kind holds (or, at worst,
such a theorem would require a mild modification of the construction).

In this paper, we propose a version of the `triplectic Darboux
theorem'. In addition to the general axioms formulated in
\cite{[BMS],[BM]}, the assumptions of the theorem involve conditions
that were not specified explicitly in the formulation
of~\cite{[BMS],[BM]}; however, they {\it are\/} valid in the canonical
coordinates of~\cite{[BM0]}, which suggests that they are rather
natural.  The conditions that we impose in order to prove the theorem
are formulated in terms of {\it marked functions\/} (or, Casimir
functions) of the two antibrackets given on the triplectic manifold.
Imposing these conditions allows us to demonstrate the existence of a
coordinate system in which one of the antibrackets becomes canonical
(just like the antibracket from~\cite{[BM0]}), while the other assumes
the canonical form on a submanifold $\cL$ of dimension one third of
the dimension of the triplectic manifold $\cM$.  This form of the
antibrackets will be referred to as {\it weakly canonical}. We
identify the obstruction to reducing the weakly canonical antibrackets
to the canonical form --- this is a matrix $e^i_\alpha$ whose entries
depend only on marked functions $\xi_{2\alpha}$ and $\xi_{1i}$ of the
antibrackets $(~,~)^1$ and  $(~,~)^2$, respectively; in fact, this matrix
relates the vector fields generated by the marked functions:  $(\xi_{2
\alpha},\;\cdot\;)^2= (-1)^{(\e(i)+1) \e(\alpha)}
  e^i_\alpha\,(\xi_{1i},\;\cdot\;)^1$.

The submanifold $\cL$ where both antibrackets become canonical plays a
crucial role in the theory also from the following point of view. The
Poisson bracket on $\cM$ (in the version given in \cite{[BM]}, which
is advantageous over the original proposal of~\cite{[BMS]}) becomes
non-degenerate on $\cL$ and, thus, makes $\cL$ into a symplectic
manifold. Then, the boundary conditions on the master-action are
imposed on some Lagrangian submanifold $\cL_0$ of~$\cL$, which can be
identified as the manifold of {\it fields\/} of the theory, with all
the antifields set to zero. Thus, (a Lagrangian submanifold of) the
symplectic submanifold is an essential ingredient of the triplectic
quantization scheme.  In fact, any symplectic leaf of the Poisson
bracket may be used as that symplectic submanifold.

The existence of an obstruction to the canonical form in a
full-dimensional neighborhood of a point in~$\cM$ raises the question
of whether some further requirements should be imposed on the
`triplectic data' or the physics of the quantized gauge system is in
some way sensitive only to the symplectic submanifold and to the form
the antibrackets take on it. This is left for the further
investigations; in this paper, we make one more small step in that
direction by classifying the odd vector fields of the form proposed
in~\cite{[BM]} that are compatible with the antibrackets.  We also
discuss how the weak canonical form of the antibrackets (and the
corresponding odd vector fields) coexist with the requirement that
there be an $Sp(2)$ action on the triplectic manifold. The $Sp(2)$
covariance on the triplectic manifold has not been discussed in much
detail in the literature, in particular the statement regarding the
$Sp(2)$ action in general coordinates was only implicit
in~\cite{[BMS]}. Here, we were not able to classify the weakly
canonical antibrackets into those which do, and those which do not,
admit an $Sp(2)$ action; however, an infinitesimal analysis suggests
that both cases can be realized, and, therefore, the requirement of
$Sp(2)$ covariance does not restrict the weakly canonical antibrackets
to the canonical ones.

\section{Basic definitions}\lvm
We begin with a brief reminder on the triplectic quantization in the
covariant approach. The role of the field-antifield space is played by
a $(4N-2k|N+2k)$-dimensional supermanifold~$\cM$. Let $\CM$ be the
algebra of functions on $\cM$.\footnote{All of our analysis is local,
which we will not stipulate explicitly any more.}  An antibracket on
$\cM$ is a bilinear mapping $(~,~) : \CM \times \CM \to \CM$ such that
$\e( (F,G) ) = \e(F) + \e(G) +1$ and
\begin{equation}\new
  \begin{array}{c}
    (F,G)=-\sign{F}{G}(G,F)\,,\\
    (F,GH)=(F,G)H+(-1)^{(\e(F)+1)\e(G)}G(F,H)\,,\\
    \sign{F}{H} (F,(G,H))+ {\rm cycle}(F,G,H)=0
    \label{jacobi}
  \end{array}
\end{equation}
for all $F,G,H \in \CM$. The pair of antibrackets $(~,~)^a$, $a=1,2$,
is called compatible if
\begin{equation}
  (~,~)=\alpha(~,~)^1+\beta (~,~)^2
\end{equation}
is an antibracket for arbitrary even constants $\alpha$ and~$\beta$.
This is equivalent to
\begin{equation}
  \label{Jacobi}
  (-1)^{(\e(F)+1)(\e(H)+1)}((F,G)^{\{a},H)^{b\}} + {\rm cycle}(F,G,H) =0\,,
\end{equation}
where the curly brackets stand for symmetrization of indices:
$C^{\{a}D^{b\}}=C^aD^b+C^bD^a$.  In the local coordinates $\Gamma^A$
on $\cM$, where we can write
\begin{equation}
  E^{aAB}=(\Gamma^A,\Gamma^B)^a , \qquad a=1,2 \,,
\end{equation}
the compatibility condition takes the form~\cite{[BLT]}
\begin{equation}
  \sign{A}{D} E^{\{aAC} \d_C E^{b\}BD}+
  {\rm cycle}(A,B,D)=0\,,
  \label{coordjacobi}
\end{equation}
The compatibility condition~\req{Jacobi} (or,~\req{coordjacobi}) is
often referred to as the symmetrized Jacobi identity.

\bigskip

In what follows, we use the notion of marked functions (Casimir
functions) of an antibracket.
\begin{dfn}\mbox{}

  \begin{enumerate}\addtolength{\parskip}{-6pt}

  \item A function $\varphi \in \CM$ is called a marked function of
    the antibracket $(~,~)$ if
    \begin{equation}
      (F,\varphi)=0
    \end{equation}
    for any $F \in \CM$.

  \item A collection $\phi_1,\ldots,\phi_n$ of marked functions of
    some antibracket $(~,~)$ is called complete if any marked function
    $\varphi$ of the antibracket is a function of only
    the~$\phi_1,\ldots,\phi_n$.
  \end{enumerate}
\end{dfn}
Thus, the marked functions from a complete set generate the algebra of
marked functions. A characteristic example is provided by the
coordinate functions that are transversal to a symplectic leaf of a
chosen (anti)bracket. In what follows, we will always take {\it
  minimal\/} complete sets (with the minimal possible number of the
$\phi_j$ functions). The number of functions in such a set is then the
co-rank of the antibracket (i.e., by definition, the codimension of
its symplectic leaf).

\bigskip

In the language of marked functions, one can observe that the
`canonical' antibrackets~\cite{[BM0]}, which can be written in the
form
\begin{equation}
  (F,G)^a=\ddr{F}{x^i}\ddl{G}{\xi_{ai}}-\sign{F}{G}(F \leftrightarrow G)\,
  \label{canonical}
\end{equation}
in some coordinate system $\Gamma^A=(x^i,\xi_{ai})$, $i=1,\ldots,2N$,
$a=1,2$, satisfy certain properties that have not been explicitly
stated before. Thus, the fact that the corresponding matrices
$E^{aAB}$ have no common zero modes, means that the only marked
functions shared by the two antibrackets are constants --- which we
will express by saying that the antibrackets do not have common marked
functions. Such antibrackets will be called {\it jointly
  nondegenerate\/}.  Further, the coordinates $\xi_{1i}$ make up a
complete set of marked functions of the second ($a=2$) antibracket
from~\req{canonical}, while $\xi_{2i}$ are a complete set of marked
functions of the first ($a=1$) antibracket. Thus, if $\phi_1$ and
$\psi_1$ (respectively, $\phi_2$ and $\psi_2$) are any two marked
functions of the second (resp., the first) antibracket, then
\begin{equation}
  (\phi_1,\psi_1)^1=0\,,    \qquad    (\phi_2,\psi_2)^2=0 \,.
  \label{flatcond1}
\end{equation}
The antibrackets whose marked functions satisfy~\req{flatcond1} will be
called {\it mutually flat}.  An important point is that this
condition, being formulated in terms of marked functions of two
antibrackets, is coordinate-independent.  Observe that it is not
necessarily fulfilled for marked functions of two compatible
antibrackets.

The following fact is a consequence of some simple linear algebra.
\begin{fact}
  Let two antibrackets be jointly nondegenerate and mutually flat.
  Then their ranks $r_a$, $a=1,2$, satisfy \ $r_a\geq3N$ \ and \
  $r_1+r_2\geq8N$, where the dimension of the manifold is
  $\dim\cM=6N$.
\end{fact}

The triplectic antibrackets, therefore, minimize the quantity
$r_1+r_2$ among all the jointly nondegenerate and mutually flat
antibrackets.

\section{Finding weak canonical coordinates\label{sec:weak}}\lvm
Given two compatible antibrackets that are jointly nondegenerate and
mutually flat, we are going to simplify them by choosing an
appropriate coordinate system on $\cM$. As in the above, the
triplectic manifold $\cM$ is of dimension $6N$ and each of the
antibrackets is of rank~$4N$.  The mutual flatness condition means
that
\begin{equation}
  (\xi_{1i},\xi_{1j})^1=0\,, \quad (\xi_{2\alpha},\xi_{2\beta})^2=0\,,
  \qquad
  i,j=1,\ldots,N\,,\quad\alpha,\beta=1,\ldots,N\,,
  \label{flatcond}
\end{equation}
where $\xi_{1i}$ is a full set of marked functions of the second
antibracket and $\xi_{2\alpha}$, those of the first one. By virtue of
the assumptions made, there are no common marked functions of the two
antibrackets; and, moreover, there exist functions~$x^i$,
$i=1,\ldots,N$, such that $(x^i,\xi_{1i},\xi_{2 \alpha})$ is a local
coordinate system on~$\cM$.

We first show that $x^i$ can be chosen in such a way that
$(x^i,\xi_{1j})^1=\delta^i_j$.  Indeed, vector fields
\begin{equation}
  X^1_i=-(\xi_{1i},~\cdot~)^1
  \label{X1field}
\end{equation}
are linearly independent at every point (because the antibrackets are
jointly nondegenerate) and, moreover, the Jacobi identity for
$(~,~)^1$ combined with the first of Eqs.~\req{flatcond} show that
these vector fields pairwise commute.  Let $\cL$ be an integral
manifold of the $X^1_i$.  Making use of the Frobenius theorem for
supermanifolds~\cite{[SHANDER]}, we construct a coordinate system
$x^i,y_A$, $A=1,\ldots,4N$, on $\cM$ in which $X^1_i=\ddl{}{x^i}$
(thus, $\cL$ is singled out by the equations $y_A=\const$).  Then
Eqs.~\req{flatcond} imply
\begin{equation}
  \ddl{}{x^i} \xi_{1j}=
  \ddl{}{x^i} \xi_{2\alpha}=0\,,
\end{equation}
therefore $\xi_{a}=\xi_{a}(y_A)$, where
$\xi_a=(\xi_{1i},\xi_{2\alpha})$.  Once all of the functions~$\xi_{a}$
are independent, we can go over to the coordinate system
$(x^i,\,\xi_{1j},\,\xi_{2 \alpha})$, where we have
$(x^i,\xi_{1j})^1=\delta^i_j$. Hence, in particular, the Grassmann
parities are
\begin{equation}
  \epsilon(x^i)\equiv\epsilon(i)\,,\quad\epsilon(\xi_{1i})=\epsilon(i)+1\,.
\end{equation}
For the future use, denote also
\begin{equation}
  \epsilon(\xi_{2\alpha})\equiv\epsilon(\alpha)+1\,.
\end{equation}

\medskip

Having achieved $(x^i,\xi_{1j})^1=\delta^i_j$, we can simplify the
antibrackets $(~,~)^1$ and $(~,~)^2$ further. Let us look at the
analogue of \req{X1field} for the second antibracket:
\begin{equation}
  X^2_\alpha=-(\xi_{2 \alpha},~.~)^2=
  (-1)^{\epsilon(\alpha)(\epsilon(i)+1)} e^i_\alpha \ddl{}{x^i}
  +A_{\alpha j} \ddl{}{\xi_{1j}}+
  A_{\alpha \beta}\ddl{}{\xi_{2 \beta}} \,,
  \label{X2field}
\end{equation}
where we have written the general form involving the
$A_{\alpha\,\cdot}$ coefficients. The Grassmann parities are
$\epsilon(e^i_\alpha)=\epsilon(i) + \epsilon(\alpha)$. Now,
\begin{equation}\new
  \begin{array}{rclclcl}
    A_{\alpha j}&=&X^2_\alpha \xi_{1j}&=&-(\xi_{2 \alpha},\xi_{1j})^2&=&0\,,\\
    A_{\alpha \beta}&=&X^2_\alpha \xi_{2 \beta}&=
    &-(\xi_{2 \alpha},\xi_{2 \beta})^2&=&0\,.
  \end{array}
\end{equation}
Therefore, the above submanifold $\cL\subset\cM$ is at the same time
an integral manifold of the vector fields $X^2_\alpha$.

In addition, it follows from Eqs.~\req{flatcond} that
\begin{equation}
  \left[X^1_i,X^2_\alpha \right]=X^1_i X^2_\alpha -
  (-1)^{\epsilon(i) \epsilon(\alpha)}
  X^2_\alpha X^1_i=0\,
\end{equation}
and therefore the functions $e^i_\alpha$ depend only on
$\xi_{1i}$, and $\xi_{2\alpha}$.

Denote
\begin{equation}
  \eta^{ij}=(x^i,x^j)^1\,.
\end{equation}
By virtue of the symmetrized Jacobi identities, $\eta^{ij}$ depend
only on $\xi_{1i}$, and $\xi_{2\alpha}$ and satisfy the equations
\begin{equation}
  \sign{i}{k} \ddl{}{\xi_{1i}}\eta^{jk}+ {\rm cycle}(i,j,k)=0\,,
  \label{conscond1}
\end{equation}
whence
\begin{equation}
  \eta^{ij}(\xi_1,\xi_2)=\ddl{}{\xi_{1i}}f^j(\xi_1,\xi_2) -
  \sign{i}{j} \ddl{}{\xi_{1j}}f^i(\xi_1,\xi_2)
\end{equation}
for some $f^i$ (which, again, are functions of only $\xi_{1i}$ and
$\xi_{2\alpha}$).  Observe that $f^i$ are defined up to the
arbitrariness of the form
\begin{equation}
  f^i(\xi_{1},\xi_{2}) \to f^i(\xi_{1},\xi_{2})+
  \ddl{}{\xi_{1i}}H(\xi_{1},\xi_{2})\,.
  \label{freedom}
\end{equation}

The functions $f^i$ can be used to define new coordinates ${\tilde
  x}^i = x^i-f^i(\xi_1,\xi_2)$, in which $({\tilde x}^i,{\tilde
  x}^j)^1=0$. At the same time, antibrackets with the $\xi_a$ do not
change: $({\tilde x}^i,\xi_{1j})^b=(x^i,\xi_{1j})^b$ and
$({\tilde x}^i,\xi_{2\alpha})^b=(x^i,\xi_{2\alpha})^b$.

Thus, we have found a local coordinate system $(x^i,\xi_{1j},\xi_{2
  \alpha})$ in which (removing the tilde)
\begin{equation}\new
  \begin{array}{rclcrcl}
    (x^i,\xi_{1j})^1&=&\delta^i_j\,,&&(x^i,\xi_{2 \alpha})^1&=&e^i_\alpha\,,\\
    (x^i,x^j)^1&=&0\,,&&(x^i,x^j)^2&=&\lambda^{ij}\,,
  \end{array}
  \label{midlevid}
\end{equation}
with all the other pairwise antibrackets of the coordinate functions
vanishing.  The symmetrized Jacobi identities for the antibrackets of
the form~\req{midlevid} show that the functions $\lambda^{ij}$ depend
only on $\xi_{1i},\xi_{2\alpha}$ and satisfy the equations
\begin{equation}\new
  \begin{array}{rcl}
    \sign{i}{k} \ddl{}{\xi_{1i}} \lambda^{jk} + {\rm cycle}(i,j,k)&=&0\,,\\
    \sign{i}{k} e^i_\alpha \ddl{}{\xi_{2 \alpha}}\lambda^{jk}+
    {\rm cycle}(i,j,k)&=&0\,,
  \end{array}
  \label{conscond23}
\end{equation}
while matrix $e^i_\alpha$ satisfies
\begin{equation}\new
  \begin{array}{rcl}
    \ddl{}{\xi_{1i}}e^j_\alpha-
    \sign{i}{j} \ddl{}{\xi_{1j}}e^i_\alpha&=&0\,,\\
    e^i_\alpha \ddl{}{\xi_{2\alpha}}e^j_\beta-
    \sign{i}{j} e^j_\alpha \ddl{}{\xi_{2\alpha}}e^i_\beta&=&0\,.
  \end{array}
  \label{econd}
\end{equation}

We now use the freedom~\req{freedom} in the definition of $x^i$ in
order to make $\lambda^{ij}(\xi_{1},\xi_{2})$ vanish. Changing
the coordinates as
\begin{equation}
  x^i\mapsto x^i-\ddl{}{\xi_{1i}}H(\xi_1,\xi_2)\,,
\end{equation}
we would have $(x^i,x^j)^2=0$ whenever $H(\xi_1,\xi_2)$ satisfies the
equations
\begin{equation}
  e^i_\alpha \ddl{}{\xi_{2 \alpha}} \ddl{}{\xi_{1j}}H
  -\sign{i}{j}
  e^j_\alpha \ddl{}{\xi_{2 \alpha}} \ddl{}{\xi_{1i}}H=\lambda^{ij}\,.
  \label{Tyutineq}
\end{equation}
Compatibility conditions for \req{Tyutineq} are satisfied by virtue
of~\req{conscond23} and~\req{econd}. We, thus, assume the solution
to~\req{Tyutineq} to exist.

\medskip

To summarize, we have arrived at
\begin{thm}
  For compatible rank-$4N$ antibrackets that are mutually flat and
  jointly nondegenerate on the triplectic manifold, there exists a
  coordinate system in which the antibrackets take the form
  \begin{equation}\new
    \begin{array}{rcl}
      (F,G)^1&=&\ddr{F}{x^i}\ddl{G}{\xi_{1i}}-
      \sign{F}{G}(F \leftrightarrow G)\,,\\
      (F,G)^2&=&\ddr{F}{x^i} \,e^i_\alpha\, \ddl{G}{\xi_{2 \alpha}}-
      \sign{F}{G}(F \leftrightarrow G)\,,
    \end{array}
    \label{weakcanonicalform}
  \end{equation}
  where the functions $e^i_\alpha=e^i_\alpha(\xi_1,\xi_2)$ make up a
  nondegenerate square matrix and satisfy Eqs.~\req{econd}.  This form
  of the antibrackets will be called {\sl weakly canonical}.
\end{thm}
The nondegeneracy of $e^i_\alpha$ follows from the rank assumptions.
Note also that the symmetrized Jacobi identities for antibrackets
\req{weakcanonicalform} are equivalent to equation~\req{econd}.  The
functions $e^i_\alpha$ are, in general, an obstruction to transforming
the triplectic antibrackets to the canonical form of~\cite{[BM0]}.

A more invariant way to look at the $e^i_\alpha$ is to consider them
as a matrix relating vector fields~\req{X1field} and \req{X2field}:
\begin{equation}
  (\xi_{2 \alpha},\;\cdot\;)^2=
  (-1)^{(\e(i)+1) \e(\alpha)}
  e^i_\alpha\,(\xi_{1i},\;\cdot\;)^1\,.
  \label{Xconnection}
\end{equation}
It follows from~\req{Xconnection} that under a change of marked
functions $\xi_{1i}\mapsto\theta_{1i}(\xi_1)$, $\xi_{2\alpha}\mapsto
\theta_{2\alpha}(\xi_2)$, the quantities $e^i_\alpha$ transform as
\begin{equation}\new
  \begin{array}{c}
    (\theta_{2 \alpha},\;\cdot\;)^2 =
    (-1)^{(\e(i)+1) \e(\alpha)}
    \tilde e^i_\alpha\,(\theta_{1i},\;\cdot\;)^1,\quad
    \tilde e^i_\alpha=
    \dl{ \xi_{1 j} }{ \theta_{1 i} }
    e^j_\beta
    \dl{\theta_{2 \alpha}}{\xi_{2 \beta}} \,.
    \label{etransform}
  \end{array}
\end{equation}
In this sense, the structure $e^i_\alpha$ depends only on the marked
functions allowed by the antibrackets. An important consequence of
\req{etransform} is that once $e^i_\alpha$ take the form
$e^i_\alpha=\delta^i_\alpha$ for some set of marked functions, then
any other choice of marked functions would leave $e^i_\alpha$ in the
class of function of the form
\begin{equation}
  e^i_\alpha(\xi_1,\xi_2)=
  e^{(1)i}_j(\xi_1) \delta^j_\beta e^{(2) \beta}_\alpha(\xi_2)\,.
  \label{reducible}
\end{equation}
This motivates the following definition.
\begin{dfn}  \label{def:reducible}
  The structure $e^i_\alpha$ is called reducible if it can be
  represented in the form~\req{reducible}
  where the functions $e^{(1)}$ and $e^{(2)}$ depend only on
  $\xi_1$ and $\xi_2$ respectively.
\end{dfn}

Conversely, whenever the matrix $e^i_\alpha(\xi_1,\xi_2)$ is reducible,
Eqs.~\req{econd} imply that there exist nondegenerate mappings
$\xi_{1i}\mapsto\theta_{1i}(\xi_1)$ and
$\xi_{2\alpha}\mapsto\theta_{2\alpha} (\xi_2)$ such that
\begin{equation}
  e^{(1)i}_j=\ddl{\theta_{1j}}{\xi_{1i}} \qquad
  e^{(2) \beta}_\alpha=\ddl{\xi_{2\alpha}}{\theta_{2\beta}} \,,
\end{equation}
choosing which as the new bases of marked functions we bring the
$e^i_\alpha$ matrix to the form $e^i_\alpha=\delta^i_\alpha$. We
conclude that the condition that $e^i_\alpha$ be reducible is
sufficient for the existence of canonical coordinates. However, it
remains a problem to formulate the property of antibrackets (and/or
their marked functions) on a triplectic manifold that would imply
reducibility.

\section{Odd vector fields and the Poisson bracket}\lvm
In addition to the antibrackets subjected to the compatibility
condition, the triplectic quantization formalism involves two odd
vector fields~$V^a$, $a=1,2$, (see~\cite{[BM],[BMS],[DN]}) that are
required to differentiate the antibrackets in following sense:
\begin{equation}
  V^{\{a}(F,G)^{b\}}=(V^{\{a}F,G)^{b\}} + (-1)^{\e(F)+1}(F,V^{\{a}G)^{b\}}
  \label{tridiff}
\end{equation}
for any two functions~$F,G \in \CM$.  Further, $V^a$ must obey the
condition
\begin{equation}
  V^{ \{ a } V^{b\}}=0\,.
  \label{nilpotency}
\end{equation}

In local coordinates $\Gamma^A$ on $\cM$, we write $V^a=(-1)^{\e(A)}
V^{aA} \d_A$.  Following~\cite{[BM]}, we restrict ourselves to the
$V^a$ fields of the following form:
\begin{equation}
  V^a=(-1)^{\e(C)} E^{aCB} F_B \d_C = (-1)^{\e(B)} F_B E^{aBC} \d_C\,,
  \qquad\e(F_A)=\e(\Gamma^A)\,,
  \label{triVvid}
\end{equation}
with some covector field $F=F_A d\Gamma^A$.
Then Eq.~\req{tridiff} rewrites as~\cite{[BM]}
\begin{equation}
 E^{ \{a AC } F_{CD} (-1)^{\e(D)}  E^{b\} DB}=0 \,, \qquad F_{AB}=
 \d_A F_B  -  (-1)^{ \e(A)\e(B)}  \d_B F_A \,.
\label{coorddiff}
\end{equation}

We now investigate the structure of these vector fields (i.e., of the
constraints~\req{coorddiff}) in the coordinates $(x^i,\xi_{1i},\xi_
{2\alpha})$ from the previous section.  The respective components of
$F$ are then denoted as $F=(F_i,F^{1i},F^{2 \alpha})$.  Using the fact
that the matrix $e^i_\alpha$ is invertible, we conclude from
Eqs.~\req{coorddiff} that there exist functions $H^1$ and $H^2$ such
that
\begin{equation}
  F_i=\ddl{}{x^i}H^{(1)}=\ddl{}{x^i}H^{(2)}\,,\qquad
  F^{1i}=\ddl{}{\xi_{1i}}H^{(1)}\,,\qquad
  F^{2 \alpha}=\ddl{}{\xi_{2 \alpha}}H^{(2)}
  \label{Fform}
\end{equation}
and the function $H=H^{(2)}-H^{(1)}$ is independent of the $x^i$
coordinates:
\begin{equation}
  \ddl{}{x^i}H=\ddl{}{x^i}(H^{(2)}-H^{(1)})=0\,,
\end{equation}
while its $\xi$-dependence is governed by
\begin{equation}
  e^i_\alpha(\xi_1,\xi_2) \ddl{}{\xi_{2 \alpha}}
  \ddl{}{\xi_{1j}}H(\xi_1,\xi_2)
  -\sign{i}{j}
  e^j_\alpha(\xi_1,\xi_2) \ddl{}{\xi_{2 \alpha}}
  \ddl{}{\xi_{1i}}H(\xi_1,\xi_2)=0\,.
  \label{Tyutineq2}
\end{equation}
This can be reformulated as follows.
\begin{fact}
  Let a pair of compatible antibrackets be written in the
  form~\req{weakcanonicalform}, and the vector fields $V^a$ represented
  as in~\req{triVvid} in some local coordinates~$x^i,\xi_{1i},\xi_{2
    \alpha}$ on~$\cM$. The vector fields differentiate the
  antibrackets if and only if there exist functions $H(\xi_{1i},\xi_{2
    \alpha})$ and $h(x^i,\xi_{1i},\xi_{2 \alpha})$ such that
  \begin{equation}\new
    \begin{array}{rcl}
      V^1&=&(-\half H,~.~)^1+(h,~.~)^1\\
      V^2&=&(\half H,~.~)^2+(h,~.~)^2\,.
    \end{array}
    \label{Vform}
  \end{equation}
  and the function $H$ satisfies Eq.~\req{Tyutineq2}.
\end{fact}

Thus, two vector fields that differentiate the antibrackets can be
split into a `Hamiltonian' or, symmetric, part $V^a_{\rm S}=(h,\cdot)^a$
and an `anti-Hamiltonian' (antisymmetric) part
$V^1_{\rm A}=(-\half H,\cdot)^1$, $V^2_{\rm A}=(\half H,\cdot)^2$.
It is easy to see that functions $H$ and $h$ are defined up to the
following transformation:
\begin{equation}\new
\begin{array}{rcl}
H & \to & H+Q^1(\xi_1)+Q^2(\xi_2) \,, \\
h & \to & h+ \half Q^1(\xi_1) - \half Q^2(\xi_2)\,,
\end{array}
\label{Harbitrariness}
\end{equation}
where functions $Q^1$ and $Q^2$ depend only on $\xi_1$ and $\xi_2$
respectively. This arbitrariness, which we will need later, follows
from the existence of vector fields $V^a$ that can be represented in
the symmetric as well as antisymmetric form.

Note also that the `anti-Hamiltonian' vector fields $V^a_{\rm A}$
differentiate the antibrackets even before symmetrizing with respect
to the $a,b$ indices,
\begin{equation}
    V^a_{\rm A}(F,G)^b=(V^a_{\rm A} F,G)^b + (-1)^{\e(F)+1}(F,V^a_{\rm
    A} G)^b\,,
\end{equation}
  which is the property postulated in~\cite{[BM0],[BMS]}.

\medskip

For the anti-Hamiltonian $V^a_A$, which we choose from now on, it is
easy to see that Eq.~\req{nilpotency} holds automatically, since this
amounts to the equation $(H,H)^a=0$, which is satisfied by
$H(\xi_{1},\xi_{2})$.\footnote{For the general form \req{Vform}, on the
  other hand, one has to take the `Hamiltonian' $h$ such that $\half
  (h,h)^a - V^a_{\rm A} h=0$, as in~\cite{[BM]}.}

\bigskip

It was observed in~\cite{[BMS]} and then developed in~\cite{[BM]} that
the pair of compatible antibrackets give rise to a Poisson
bracket\footnote{See also~\cite{[DN],[SZ]}, where the Poisson
  structure was discussed in the context of the standard BV
  quantization.}. To this end~\cite{[BM]}, one defines on $\cM$ a
tensor field
\begin{equation}
  \omega^{AB}=\half \epsilon_{ab} (-1)^{\e(D)} E^{aAC} F_{CD}E^{bDB}\,.
  \label{triPBmatrix}
\end{equation}
Due to~\req{coorddiff}, this is (graded) antisymmetric,
$\omega^{AB}=-\sig{A}{B} \omega^{BA}$. Thus, we have an antisymmetric
bracket operation
\begin{equation}
  \{F,G\}=F \dpr_A \omega^{AB} \dpl_B G \,.
  \label{PB}
\end{equation}
This structure is invariant under changing $F_A$ from~\req{triVvid} to
$F_A+\d_A K$ with an arbitrary $K$.

\begin{fact}
  Under the assumptions of Sect.~\ref{sec:weak} --- i.e., for
  compatible rank-$4N$ antibrackets that are mutually flat and jointly
  nondegenerate and for the vector fields of the form \req{triVvid}
  that differentiate the antibrackets as in \req{tridiff}, ---
  $\omega^{AB}$ is a Poisson structure and the integral submanifold
  $\cL$ associated with the antibrackets contains its symplectic leaf.
\end{fact}
Indeed, in the weakly canonical coordinates $(x^i,\xi_{1i},\xi_
{2\alpha})$ from the previous section, the only nonvanishing
components of $\omega^{AB}$ are $\omega^{ij}=\{x^i,x^j\}$. In
particular, therefore, $\rank\omega^{\cdot\;\cdot}\leq2N$.
Using~\req{Fform} for the components of $F$ in the
$(x^i,\xi_{1i},\xi_{2 \alpha})$ coordinates, we arrive at
\begin{equation}
  \omega^{ij}=(-1)^{\e(j)}\,e^i_\alpha\,
  \ddl{}{\xi_{2 \alpha}} \ddl{}{\xi_{1j}}H\,.
\end{equation}
Thus, \req{PB} becomes a Poisson bracket on $\cM$ --- i.e., satisfies
the Jacobi identity --- due to the simple fact that $H$ is independent
of~$x^i$. The triplectic manifold $\cM$ is then endowed with a Poisson
structure.

Observe that, while in general $\rank\omega^{\cdot\;\cdot}\leq2N$, the
physical requirements of quantization (i.e., of the construction of
path integral) is such that $\rank\omega^{\cdot\;\cdot}=2N$, in which
case the manifold $\cL$, which in the local coordinates is defined by
$\xi_{1i}=\const_i$, $\xi_{2 \alpha}=\const_\alpha$, is a symplectic
leaf of the Poisson bracket~\req{PB}.  Vice versa, a Lagrangian
submanifold $\cL_0$ of any symplectic leaf $\cL$
of the Poisson bracket can be used in the triplectic quantization in
order to impose boundary conditions on the master-action: one
identifies this Lagrangian submanifold as the manifold of `classical' fields.

It may also be noted that we have avoided {\it imposing\/} on the
vector fields the additional constraints of~\cite{[BM]}, namely (in
terms of the `potential' $F_A$ from~\req{triVvid}), $ F_B
E^{aBC}F_C=0$ and $ F_{AB} E^{aBC} F_{CD}=0$, which are
fulfilled automatically for the `anti-Hamiltonian' part $V^a_{\rm A}$
in our approach.

\section{$Sp(2)$-covariance}\lvm
Until this moment, we have not discussed the $Sp(2)$-covariance of our
construction.  The standard formulation~\cite{[BLT]} of the
$Sp(2)$-symmetric Lagrangian quantization assumes that the $Sp(2)$
group acts on the phase space coordinates, which are in fact $Sp(2)$
tensors; the antibrackets and the odd vector fields defined in
\cite{[BLT],[BM0]} carry the $Sp(2)$ vector representation index.

We have to extend the $Sp(2)$-covariance requirement to the
geometrically covariant formulation.\footnote{We thank I.~Tyutin for a
  discussion of this point.} \ Let us note first of all that the
conditions imposed on the antibrackets (that they be compatible,
mutually flat and jointly nondegenerate and have rank $4N$, while the
vector fields of the form \req{triVvid} satisfy~\req{tridiff}
and~\req{nilpotency}) are preserved by $Sp(2)$-transformations acting
on the $a$ index of the antibrackets and the vector fields. Now, this
action has to be realized in terms of an $Sp(2)$-action on $\cM$.

Let $\phi$ be an $Sp(2)$ action on $\cM$, i.e., to every $G \in
Sp(2)$ there corresponds a mapping $\phi_G : \cM \to \cM$ such that
$\phi_{G_1}\phi_{G_2}=\phi_{G_1 G_2}$. The pullback $\phi_G^\#$ acts
on functions in the standard way:
\begin{equation}
  (\phi^\#_G(f))(p)=f(\phi_G(p))\,,\quad p\in\cM\,,
  \label{funcmap}
\end{equation}
then $\phi^\#_{G_1} \phi^\#_{G_2}=\phi^\#_{G_2 G_1}$.
\begin{dfn}\label{def:covariance}
  A pair of compatible antibrackets and odd vector fields
  $V^a$ on $\cM$ are called $Sp(2)$ covariant if for any $G \in
  Sp(2)$ there exists a mapping $\phi_G: \cM \to \cM$ such that
  \begin{equation}
    \phi^\#_G((f,g)^a)=
    G^a_b(\phi^\#_G(f),\phi^\#_G(g))^b\,,
    \qquad \phi^\#_G(V^af)=G^a_bV^b(\phi^\#_G(f))\,,
    \label{spcovariance}
  \end{equation}
  for all $f,g\in C_{\cM}$.
\end{dfn}

In the infinitesimal form, we have a mapping from the Lie algebra
$sp(2)$ to the algebra of vector fields on $\cM$. Let, in some
coordinate system, $Y=Y^A\d_A$ be the vector field corresponding to $g
\in sp(2)$. Then the $sp(2)$ covariance condition takes the following
form:
\begin{equation}
  (L_Y E^a)^{AB}=
  Y^C\d_C E^{aAB}-Y^A\dpr_C E^{aCB}-E^{aAC}\d_C Y^B
  =g^a_b E^{bAB}\,,
  \label{Ecov}
\end{equation}
\begin{equation}
  L_Y V^a=[Y,V^a]=g^a_b V^b\,,
  \label{Vcov}
\end{equation}
that is, $Y$ acts by the Lie derivative.
The last equation \req{Vcov} imposed on the vector fields
$V^a$ of the form~\req{triVvid} one can rewrite as
\begin{equation}
  L_Y F=0\,,
  \label{Fcov}
\end{equation}
where $F=F_A d\Gamma^A$ is the covector that defines the vector
fields.  It also follows from the $Sp(2)$ covariance of the
antibrackets and vector fields $V^a$ of the form \req{triVvid} that
the Poisson bracket \req{PB} is an $Sp(2)$ scalar:
\begin{equation}
  L_Y \omega=0\,.
\end{equation}

\smallskip

Now, we are interested in whether the $Sp(2)$ covariance is compatible with the
weakly canonical for of the antibrackets, i.e., whether there exist vector
fields representing the $sp(2)$ action under which the weakly canonical
antibrackets and the odd vector fields $V^a$ are covariant in the above sense.
Let $Y^\pm,\,Y^0$ be the vector fields that implement the infinitesimal action
of the respective $sp(2)$ generators $J^\pm\,J^0$, respectively.  We denote by
$(Y^i,Y_{1j},Y_{2\alpha})$ the components in the coordinates
$(x^i,\xi_{1i},\xi_{2\alpha})$, with $\epsilon(Y^i)=\epsilon(i)$,
$\epsilon(Y_{1j})=\epsilon(j) + 1$, and
$\epsilon(Y_{2\alpha})=\epsilon(\alpha)+1$.  Then, \req{Ecov} implies that
$\ddr{Y^i}{x^j}={\d Y_{1j}\over\d\xi_{1i}}$ for either $Y^+$ or $Y^-$.
Moreover, $Y_{1j}^{\pm,0}$ depend only on $\xi_a$, which allows us to express
${Y^j}^{\pm,0}$ through $Y_{1j}^{\pm,0}$, while the remaining components must
satisfy the following equations:
\begin{equation}\new
  \begin{array}{c}
    \ddl{}{\xi_{1i}} (Y^+_{1k} e^k_{\alpha})+
    Y^+_{2 \beta}\ddl{}{\xi_{2\beta}} e^i_\alpha-
    e^i_\beta  \ddl{}{\xi_{2 \beta}} Y^+_{2 \alpha} = 0\,, \\
    \ddl{}{\xi_{2 \alpha}} Y^+_{1j} = 0\,,\qquad
    \ddl{}{\xi_{1i}} Y^+_{2 \alpha} = e^i_\alpha \,,
  \end{array}
  \label{coordEcov2}
\end{equation}
and
\begin{equation}\new
  \begin{array}{c}
    \ddl{}{\xi_{1i}} (Y^-_{1k} e^k_{\alpha})+
    Y^-_{2 \beta}\ddl{}{\xi_{2\beta}} e^i_\alpha-
    e^i_\beta  \ddl{}{\xi_{2 \beta}} Y^-_{2 \alpha} = 0\,, \\
    e^i_\alpha \ddl{}{\xi_{2 \alpha}} Y^-_{1j} = \delta^i_j \,,\qquad
    \ddl{}{\xi_{1i}} Y^-_{2 \alpha} = 0
  \end{array}
  \label{coordEcov1}
\end{equation}
(and a similar equation for $Y^0$).

The problem now is whether these equations on the components of $Y$
can be satisfied for generic $e^i_\alpha(\xi_1,\xi_2)$ subjected to
Eqs.~\req{econd} or the existence of a solution implies further
restrictions on~$e^i_\alpha$. First of all, it is not difficult to
check directly that the reducible antibrackets are $Sp(2)$-covariant
in the sense of the above definitions:
\begin{fact}
  For a reducible $e^i_\alpha$, equations
  \req{coordEcov2}--\req{coordEcov1}
  admit a solution and, therefore,
  the antibrackets are $Sp(2)$-covariant on the triplectic manifold.
\end{fact}
For the general $e^i_\alpha$, on the other hand, the analysis of
Eqs.~\req{coordEcov2}--\req{coordEcov1} for the components of $Y$ is quite
complicated because the $e^i_\alpha$ matrix is not known explicitly.  In the
Appendix, we consider the $e$-structures that differ infinitesimally from a
constant, but is not necessarily reducible.  It follows that, even though
Eqs.~\req{coordEcov2}--\req{coordEcov1} would not be solved for any
$e^i_\alpha$, there exist irreducible $e^i_\alpha$ such that these equations
admit a solution and, thus, the antibrackets \req{weakcanonicalform} are
$Sp(2)$-covariant in the sense of definition~\req{def:covariance}.

\bigskip

Recall also that, in addition to the $Sp(2)$-covariance of the
antibrackets, one should ensure the $Sp(2)$-covariance of the odd
vector fields. Then, in addition
to~\req{coordEcov2}--\req{coordEcov1}, each of the components of the
$Y$ vector fields should satisfy equations of the following form:
\begin{equation}\new
  \begin{array}{c}
    \ddl{}{\xi_{1i}} (Y^+_{1k}\ddl{}{\xi_{1k}}H)+
    Y^+_{2 \beta}\ddl{}{\xi_{2 \beta}} \ddl{}{\xi_{1i}}H-
    (\ddl{}{\xi_{1i}} Y^+_{2 \beta})
    (\ddl{}{\xi_{2 \beta}} H)=0\,, \\
    \ddl{}{\xi_{2 \alpha}} (Y^+_{2 \beta} \ddl{}{\xi_{2 \beta}} H)+
    Y^+_{1k}\ddl{}{\xi_{1k}} \ddl{}{\xi_{2 \alpha}}H-
    (\ddl{}{\xi_{2 \alpha}} Y^+_{1k})
    (\ddl{}{\xi_{1k}}H)=0\,,
  \end{array}
  \label{coordFcov}
\end{equation}
and similarly for $Y^-$ and $Y^0$. Here,
we took the odd vector fields in the `anti-Hamiltonian' form
\begin{equation}
    V^1=(-\half H,~\cdot~)^1\,, \qquad  V^2=(\half H,~\cdot~)^2\,,
\label{LHform}
\end{equation}
with $H$ satisfying~\req{Tyutineq2}. Let us first show the following
fact, which describes the reducible situation:
\begin{fact}\label{fact:Hcan}
  If $e^i_\alpha=\delta^i_\alpha$, then Eqs.\req{coordFcov},
  \req{coordEcov2}--\req{coordEcov1} imply that the function $H$
  is at most bilinear in $(\xi_{1i},\xi_{2 \alpha})$.
\end{fact}
To prove this, we employ the general form of the vector fields $Y^\pm$
representing generators $J^\pm$ (i.e., the general solutions of
Eqs.~\req{coordEcov2}--\req{coordEcov1} with
$e^i_\alpha=\delta^i_\alpha$). The $Y$ of this form are then inserted
into each of Eqs.~\req{coordFcov} and the resulting equations are
solved making use of the fact that $H$ satisfies Eq.~\req{Tyutineq2}
Then, if $H$ is homogeneous in $\xi_a$, the desired statement follows
immediately, while in the case where $H$ is taken as a series in
$\xi_a$ a slightly more involved analysis shows that all of the
coefficients except the one in $C^{i\alpha}\xi_{1i}\,\xi_{2 \alpha}$
vanish as well.

In the reducible case, therefore, the $Sp(2)$-covariance condition for
the vector fields implies the following form of the $H$ function:
\begin{equation}
  H=\omega^{ij} \, \xi_{1i} \, \xi_{2j}+ T^{1i} \, \xi_{1i} + T^{2i}
  \xi_{2i}\,.
\end{equation}
Then, using the arbitrariness~\req{Harbitrariness} in the definition
of `anti-Hamiltonian' vector fields we can reduce the odd vector
fields to the form proposed in~\cite{[BMS]} (the Poisson matrix
$\omega^{ij}$ should be nondegenerate in the quantization context,
hence, by a linear transformation, it can be brought to the canonical
form). We can summarize our results as:
\begin{thm}Let there be given a pair of compatible antibrackets on
  $\cM$ and a pair of odd vector fields $V^a$ of the
  form~\req{triVvid} compatible with the antibrackets.
  Assume also that
  \begin{itemize}\addtolength{\parskip}{-6pt}
  \item
    the antibrackets and the odd vector fields are $Sp(2)$-covariant,
  \item
    the antibrackets have the minimal rank, are jointly nondegenerate
    and mutually flat,
  \item
    the $e^i_\alpha$ structure that corresponds to the antibrackets is
    reducible.
  \end{itemize}
  Then the antibrackets, the odd vector fields, and the corresponding
  Poisson bracket can be brought to the canonical form, which
  coincides with that proposed in~\cite{[BMS]} whenever the Poisson
  bracket is of maximal rank and therefore nondegenerate on $\cL$.
  Note also that in the case of the vector fields, the transformation
  to the canonical form may involve adding a purely Hamiltonian piece.
\end{thm}

\section{Concluding remarks}\lvm We have shown that the triplectic axioms
of~\cite{[BMS]}, together with the additional requirements imposed on the
marked functions of the two antibrackets, allow one to find a coordinate system
where the antibrackets take the weakly canonical form~\req{weakcanonicalform}.
In that formula, the structure $e^i_\alpha(\xi_1,\xi_2)$ considered modulo
reducible structures \req{reducible} is an obstruction to bringing both
antibrackets to the canonical form. The weakly-canonical antibrackets become
canonical on the symplectic submanifold~$\cL$ of the triplectic manifold. We
have also classified the `triplectic' vector fields $V^a$ taken in the
framework of the ansatz proposed in~\cite{[BM]} that are compatible with the
weakly canonical antibrackets. It follows that the conditions imposed on the
marked functions imply that the $V^a$ vector fields satisfy the constraints
postulated in~\cite{[BM]} and, therefore, induce a Poisson bracket on the
triplectic manifold~$\cM$.  We also have formulated the $Sp(2)$-covariance
condition for the antibrackets and the $V^a$ vector fields. For a reducible
$e^i_\alpha(\xi_1,\xi_2)$, this condition implies the canonical form of the
triplectic vector fields.

It should be recalled that the possibility to transform the
antibracket to the canonical form allows one to carry over to the
covariant formulation a number of important statements in the theory,
such as, for example, the statement regarding the existence and
uniqueness of solutions to the master-equation. It, thus, remains to
be seen whether working with the weakly-canonical antibrackets allows
one to prove the existence of the solution to the triplectic
master-equation.

\medskip

It may be remarked that the weakly canonical form of the antibrackets
is somewhat reminiscent of the `non-Abelian' antibrackets
of~\cite{[AD]}, however a significant difference is that the
non-Abelian antibracket involves some functions $u^A_i$ that depend on
the $\phi^A$ fields in such a way that the derivations $u^A_i
\ddl{}{\phi^A}$ make up a Lie algebra, while in our case the
antibracket involves functions $e^i_\alpha$ that depend only on the
marked functions of the antibrackets and all of the vector fields
$e^i_\alpha\ddl{}{\xi_{2 \alpha}}$ pairwise commute.

\paragraph{Acknowledgments} We are grateful to K.~Bering,
O.~Khudaverdyan, A.~Nersessian, I.~Tipunin and, especially, to
I.~Tyutin, for very useful discussions. We also wish to thank
I.~Batalin for illuminating discussions on a number of problems in
quantization of gauge theories. This work was supported in part by the
RFFI grant 96-01-00482 and by grant INTAS-RFBR-95-0829 from the
European Community.

\def\theequation{A.\arabic{equation}}
\section*{Appendix}\lvm
Here, we show how one can construct a solution for
the $e^i_\alpha$ structure that is not necessarily reducible.
The general analysis of equations \req{coordEcov2}--\req{coordEcov1},
which guarantee the $Sp(2)$-covariance of
antibrackets~\req{weakcanonicalform}, is very involved. In what
follows, we restrict ourselves to the case where $e^i_\alpha$ differs
from a constant only infinitesimally:
\begin{equation}
  e^i_\alpha(\xi_1,\xi_2) = \delta^i_\alpha + \varepsilon
  c^i_\alpha(\xi_1,\xi_2)\,.
  \label{einfinitesimal}
\end{equation}

Consider then, e.g., Eq.~\req{coordEcov2}. It now rewrites in the
following terms. We set
\begin{equation}
  Y^+ = {\bar Y}^+ + \varepsilon y^+.
\end{equation}
In the zeroth order in $\varepsilon$, we have
\begin{equation}
  {\bar Y}^+_{1m} = \xi_{1j}a^j_m\,,\qquad
  {\bar Y}^+_{2m} = \xi_{1m} + \xi_{2j}a^j_m
\end{equation}
with a constant matrix $a^j_m$.  In the first order in $\varepsilon$,
then, Eqs.~\req{coordEcov2} take the following form:
\begin{equation}\new
  \begin{array}{c}
    \ddl{}{\xi_{2i}}(y^+_{2m} - \xi_{1l}\ddl{y^+_{2m}}{\xi_{1l}})=
    \ddl{y^+_{1m}}{\xi_{1i}} +
    a^i_k\,c^k_m - c^i_k\,a^k_m
    + \xi_{1l}a^l_k\ddl{c^i_m}{\xi_{1k}}
    + \xi_{2l}a^l_k\ddl{c^i_m}{\xi_{2k}} \,,\\
    \ddl{y^+_{1m}}{\xi_{2i}}=0\,,\qquad \ddl{y^+_{2m}}{\xi_{1i}}=c^i_m.
  \end{array}
  \label{infJ+}
\end{equation}

As we saw in Sec.~5, the equations for $Y$ have a solution when the
$e^i_\alpha$ structure is reducible. To give an example of a solution
which is not reducible, we consider the function $c^i_m$ that is
homogeneous in $\xi_1,\xi_2$,
\begin{equation}
  c^i_m(\xi_1,\xi_2)=
  K^{i~p_{1}, \ldots , p_\alpha~q_1, \ldots, q_\beta}_m \,
  \xi_{1p_1}, \ldots , \xi_{1p_\alpha} \,
  \xi_{2q_1}, \ldots , \xi_{2q_\beta} \,.
  \label{K}
\end{equation}
Inserting $c^i_m$ of this form into \req{infJ+}, we can observe that,
in addition to functions depending {\it only\/} on $\xi_1$ or on
$\xi_2$, homogeneous solutions to that equation must depend linearly
on~$\xi_2$. In a similar way, we see from the equation ensuring the
covariance with respect to the $J^-$ generator that (again, in
addition to the `trivial' solutions) the solutions are linear
in~$\xi_1$.  Modulo the $c^i_j$ structures that correspond to
reducible $e^i_j$, we thus have the following general form of the {\it
  homogeneous\/} $c^i_j$ functions corresponding to $Sp(2)$-covariant
antibrackets:
\begin{equation}
  c^i_m(\xi_1,\xi_2)= K^{i~pq}_m\,\xi_{1p}\,\xi_{2q}\,.
  \label{K2}
\end{equation}

In this way, we obtain a vector field on the triplectic manifold which
ensures the $Sp(2)$ covariance of the weakly canonical, but not
reducible, antibrackets of the form~\req{weakcanonicalform} in which
$e^i_\alpha$ has the form~\req{einfinitesimal},~\req{K2}.  It can also
be shown that there exist $Sp(2)$-covariant vector fields $V^a$ that
differentiate the antibrackets defined by the above~$c^i_j$.


\small

\begin{thebibliography}{33}
  \parindent=0pt \parskip=-2pt

\bibitem{[BM0]} I.~A.~Batalin and R.~Marnelius, \PLB{446}~(1995) 44.

\bibitem{[BMS]} I.~A.~Batalin, R.~Marnelius, and A.~M.~Semikhatov,
  \NPB{446}~(1995) 249.

\bibitem{[BM]} I.~A.~Batalin and R.~Marnelius, \NPB{465}~(1996) 521.

\bibitem{[Hull]}C.M.~Hull, \MPLA5 (1990) 1871.

\bibitem{[BLT]} I.~A.~Batalin, P.~M.~Lavrov, and I.~V.~Tyutin,
  \JMP{}~{31}~(1990) 1487.\\
  I.~A.~Batalin, P.~M.~Lavrov, and I.~V.~Tyutin, \JMP~{32}~(1990) 532.\\
  I.~A.~Batalin, P.~M.~Lavrov, and I.~V.~Tyutin, \JMP~{32}~(1991)
  2513.

\bibitem{[DN]} P.~H.~Damgaard and A.~Nersessian, \PLB{355}~(1995)~150

\bibitem{[BV]} I.~A.~Batalin and G.~A.~Vilkovisky, \PLB~Vol.102~(1981) 27.\\
  I.~A.~Batalin and G.~A.~Vilkovisky, \PRD~Vol.28~(1983) 2567.

\bibitem{[BT]} I.~A.~Batalin and I~.V~Tyutin, Int. J.
  Mod. Phys.~A8~(1993)~2333.\\
  I.~A.~Batalin and I~.V~Tyutin, Mod. Phys.  Lett.~A8~(1993)~3673.\\
  I.~A.~Batalin and I~.V~Tyutin, Mod. Phys.  Lett.~A9~(1994)~1707.

\bibitem{[HZ]} H.~Hata and B.~Zwiebach, Ann. Phys. 229~(1994)~177.

\bibitem{[SZ]} A.~Sen and B.~Zwiebach, \CMP{177}~(1996)~305.

\bibitem{[SHANDER]} V.~N.~Shander, {\it Analogues of the Frobenius and
    Darboux Theorems for Supermaifold\/}, Comptes Rendus de l'Academie
  Bulgare des Sciences, Vol.36, n.3, p.309, (1983).

\bibitem{[DJB]} P.~H.~Damgaard, F.~De~Jonghe, and K.~Bering,
  \NPB{455}~(1995)~440.

\bibitem{[AD]} J.~Alfaro and P.~H.~Damgaard, \PLB{369}~(1996)~289-294.

\bibitem{[ASS]} A.~Schwarz, \CMP{155}~(1993)~249-260.

\end{thebibliography}
\end{document}